\documentclass[twocolumn,showpacs,preprintnumbers,amsmath,amssymb,prl]{revtex4-1}
\usepackage{graphicx}
\usepackage{amsmath, amsthm, amssymb}
\usepackage{latexsym}
\usepackage{amssymb}
\usepackage{bm}
\usepackage{dcolumn}
\usepackage{epstopdf}
\usepackage{color}

\newcommand{\et}{{\it et\  al.}}

\begin{document}

\title{Coherent Charge and Spin Density Waves in Underdoped HgBa$_{2}$CuO$_{4+\delta}$}

\author{Jeongseop A. Lee$^{1}$, Yizhou Xin$^{1}$, W. P. Halperin$^{1}$, A. P. Reyes$^{2}$ P. L. Kuhns$^{2}$, M. K. Chan$^{3}$}

\affiliation{$^1$Department of Physics and Astronomy Northwestern University, Evanston, Illinois 60208, USA\\$^2$National High Magnetic Field Laboratory, Tallahassee, Florida 32310, USA\\$^3$Pulsed Field Facility, National High Magnetic Field Laboratory, Los Alamos National Laboratory, Los Alamos, New Mexico 87545, USA}
\date{Version \today}

\begin{abstract}

\end{abstract} 

\pacs{ }

\maketitle

Various forms of spin and charge ordering have been identified in a wide range of cuprate superconducting materials, but whether these behaviors are ubiquitous phenomena is not established. In this work we focus on one of the simplest  compounds, HgBa$_{2}$CuO$_{4+\delta}$ (Hg1201), a superconductor with a high transition temperature, 97 K, having only a single layer and  tetragonal structure, in contrast to one of the most extensively studied materials, YBa$_{2}$Cu$_{3}$O$_{6+y}$ (Y123)~\cite{Wu2011, Chang2012, Ghiringhelli2012, Wu2013, Gerber2015}.  Using nuclear magnetic resonance  we have discovered  a  coherent spatial modulation of both spin and charge  that is temperature and magnetic field independent, in competition with superconductivity similar to other cuprates~\cite{Comin2015}.  However, there is no evidence for the magnetic field and temperature induced  charge order observed in Y123~\cite{Wu2011,Gerber2015}.  Electronic instabilities are a common feature of cuprates~\cite{Keimer2015} as in the present work on Hg1201, but their manifestations are not universal.

\begin{figure}[!ht]
\centerline{\includegraphics[width=8.5cm]{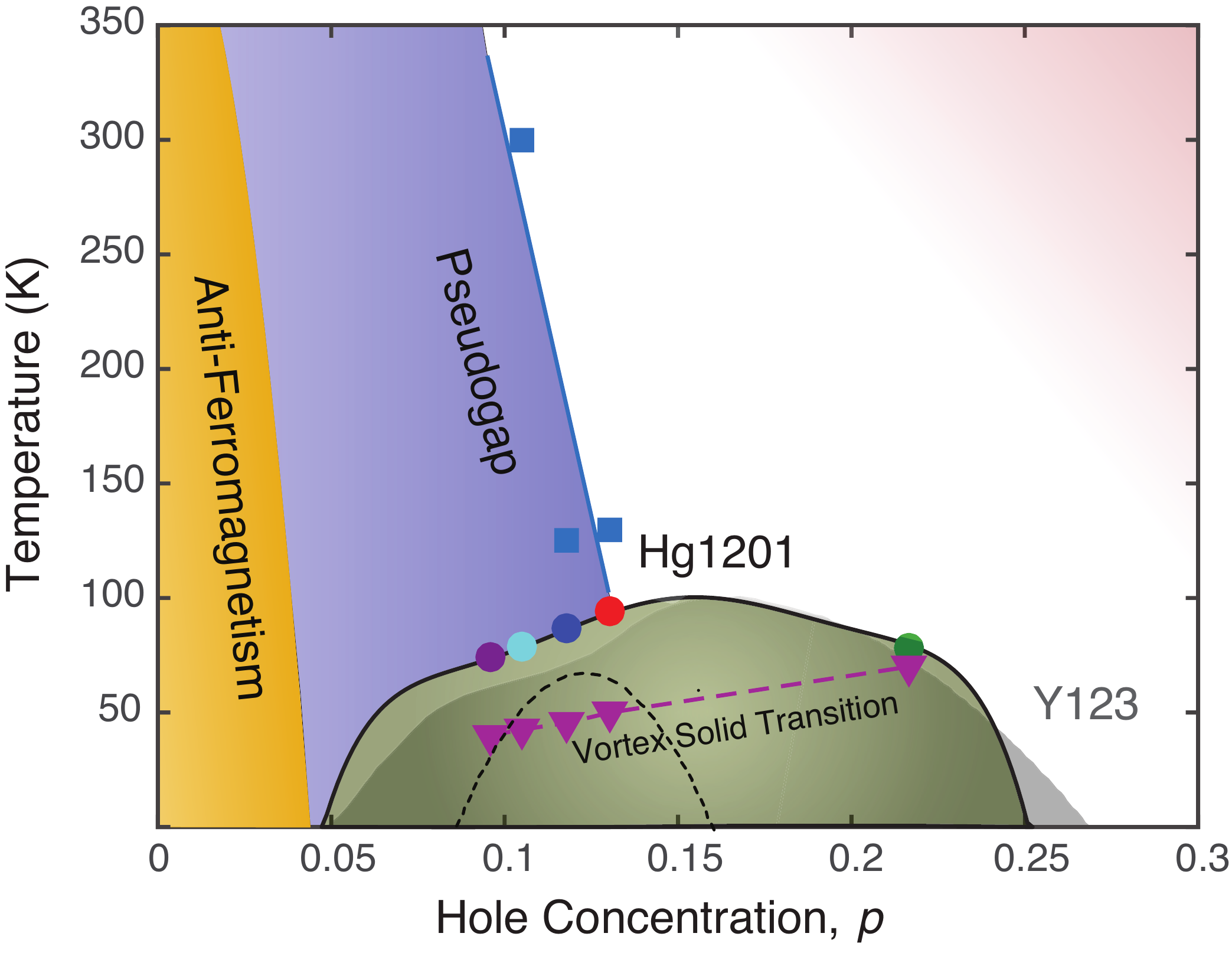}}
\caption {Phase diagram for Hg1201 taken in part from Li {\it et al.}~\cite{Li2011}.    Vortex melting and pseudogap temperatures were determined from  $^{63}$Cu and  $^{17}$O NMR~\cite{Lee2016}.  NMR results discussed in this report are for crystals (from left to right) UD74, UD79, and UD87 indicated by solid circles, together with two additional samples UD94 and OD89. The shaded dome in the background is for Y123 juxtaposed with a dashed parabolic curve indicating change in temperature dependence of the Hall coefficient.~\cite{Doiron2013,Taillefer2009}}
\label{Lee.Figure1}
\end{figure}

Efforts to better understand cuprate superconductors have focused on the nature of the pseudogap at  temperatures above the superconducting transition and the role of spatial modulations of charge or spin~\cite{Wu2011, Chang2012, Ghiringhelli2012, Wu2013, Tabis2014,Campi2015,Gerber2015,Keimer2015}.  A notable example is  high-quality, underdoped crystals of YBa$_{2}$Cu$_{3}$O$_{6+y}$(Y123)\cite{Comin2015}.  Magnetic  field and temperature induced  charge order has been reported from NMR measurements\cite{Wu2011,Wu2013}, recently confirmed as incommensurate by x-ray diffraction~\cite{Gerber2015} in Y123 near hole-doping $p=0.125$.  This behavior can be correlated with  change in sign of the temperature dependence in the Hall coefficient~\cite{Taillefer2009, Doiron2013}, shown in the phase diagram Fig.~\ref{Lee.Figure1}, that has been associated with reconstruction of the Fermi surface with electron pockets identified from quantum oscillations~\cite{Doiron2007,Barisic2013}.  

We use $^{17}$O  NMR to investigate the underdoped high temperature cuprate  HgBa$_{2}$CuO$_{4+\delta}$ (Hg1201); a tetragonal, single layer compound with optimal superconducting transition of 97\,K, arguably the simplest known high temperature superconductor.  Similar to earlier NMR results on underdoped Y123~\cite{Wu2011,Wu2013}, we have found  charge and spin density wave order in Hg1201.  However, distinct from the work in Y123 this order has a rather different character.  It is temperature and magnetic field independent and it is not induced by either.  This wide range of behaviors suggests that electronic ordering and superconductivity may not be intimately connected. 

Crystals of near optimally doped Hg1201 were  grown at the University of Minnesota.  Isotope exchange for $^{17}$O  NMR was  performed at Northwestern University followed by annealing for typically one week  to establish  doping and homogeneity.  Finally characterization was performed using $^{17}$O and $^{199}$Hg  nuclear magnetic resonance (NMR), Laue x-ray diffraction, and low field SQUID measurements.  This report is mainly concerned with two underdoped single crystals of Hg1201 with T$_{c}$ = 87\,K (UD87) and 79\,K (UD79), but with some recent work on the sample 74\,K (UD74) which we  had reported upon previously~\cite{Mounce2013}.   Their  oxygen concentrations correspond  to hole doping of $p = 0.118$,  0.105, and 0.095 respectively, obtained by comparing the measured T$_{c}$ with the phase diagram, Fig.\ref{Lee.Figure1}~\cite{Li2011}.  With perfect alignment of the crystal $c$-axis to the external field, there exists two sets of five $^{17}$O NMR spectral peaks: one set for oxygen in the CuO$_2$ plane O(1), and the other for the apical oxygen O(2).  Since the nuclear spin is $I=5/2$, these are associated with the central transition, ($\frac{1}{2}$,-$\frac{1}{2}$), and four quadrupolar satellites corresponding to the transitions, ($\frac{5}{2}$,$\frac{3}{2}$), ($\frac{3}{2}$,$\frac{1}{2}$), (-$\frac{1}{2}$,-$\frac{3}{2}$), and (-$\frac{3}{2}$,-$\frac{5}{2}$). In general the transition frequencies are labeled by indices ($m, m-1$).  For this investigation we have varied the magnetic field from $H_0 =6$ to 30\,T over a  range of temperature, from $\sim$\,4\,K to 400\,K.  

The situation where the magnetic field is not aligned  with the $c$-axis is not D4 symmetric for O(1),  resulting in anisotropic broadening or splitting of each of the peaks of the NMR spectrum since the two planar oxygens have orthogonal Cu-O bonding directions and are no longer degenerate. Therefore the O(1) rotation pattern consists of two sets of five peaks, one each for O(1a) and O(1b) where a and b denote one of two inequivalent oxygen sites in the cuprate plaquette identified by their differing projections of the magnetic field along the Cu-O bond. In contrast all apical sites O(2) are equivalent and their spectra always have exactly five peaks. This situation is very different from oxygen chain-ordered Y123~\cite{Wu2011}.

\begin{figure}[!ht]
\centerline{\includegraphics[width=8.5cm]{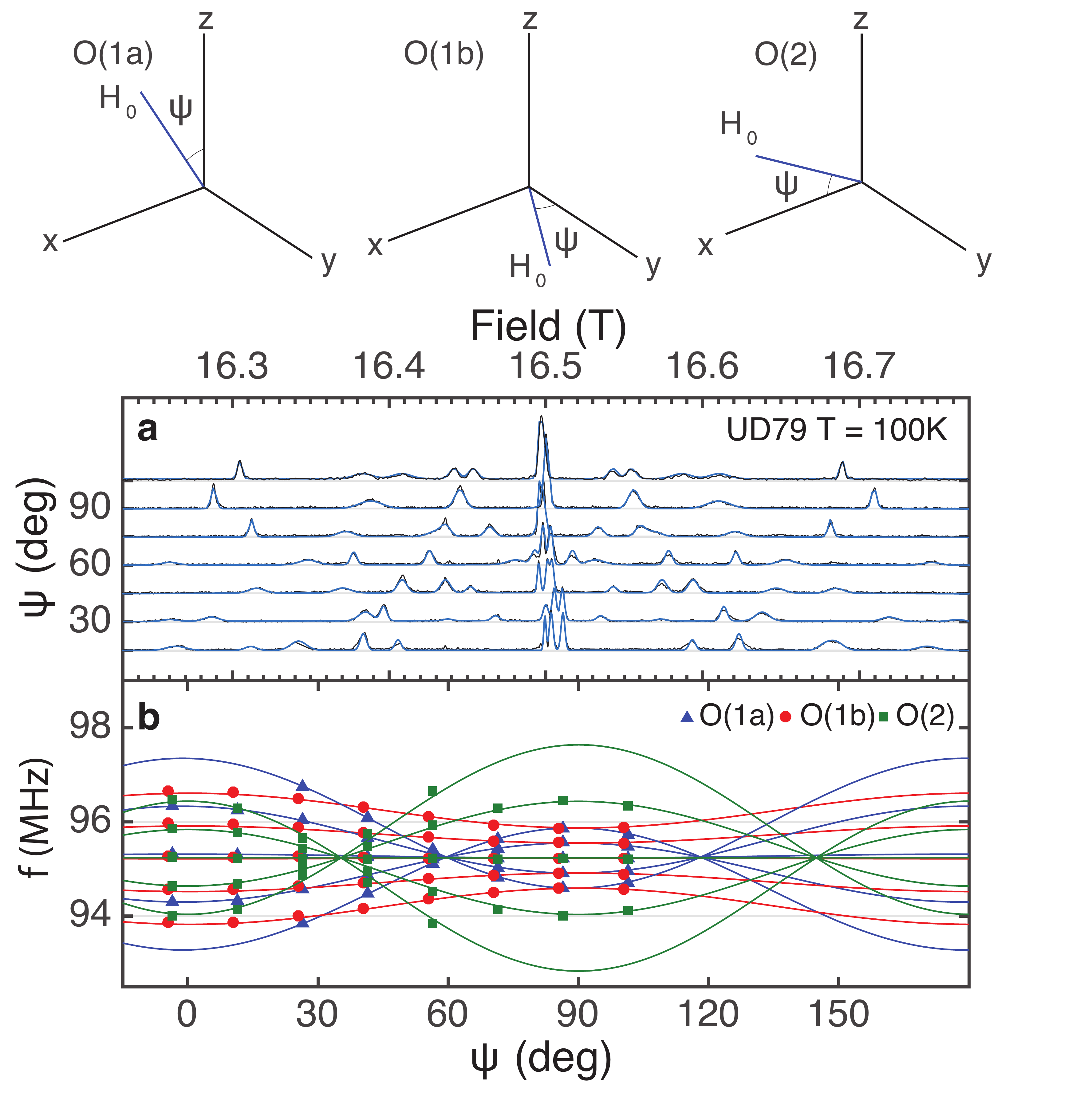}}
\caption {[a] NMR field swept spectra of  UD79 at multiple orientations at $\it{f}$  = 95.39 MHz, $H_{0}$ = 16.5\,T, and $T = 80$\,K with gaussian fits over each peak in blue lines with rotation about the b-axis offset by the goniometer  angle between $H_0$ and the $a$-axis.  [b] Peak frequency positions of spectra for O(1a), O(1b), and O(2) as a function of $\psi$. Solid curves are fits described in the text. }
\label{Lee.Figure2}
\end{figure}

Additionally, the quadrupolar NMR spectra in a magnetic field will be modified in the presence of electronic charge or spin order.  In the case of a change in the electronic spin susceptibility, all five spectral peaks of a given nucleus are affected identically and have the same shift in frequency.  In contrast, charge order produces a modulation of the local electric field gradient (EFG) which changes the quadrupolar satellite frequencies shifting them mirror-like with respect to the central peak.  In short, spin or charge order can be independently identified from their distinct  signatures in the NMR spectrum.

Given the various possible contributions to frequency shifts  it is clear that to perform an NMR investigation of charge or spin ordering one must begin  with a full determination of the rotation pattern  as shown in Fig.~\ref {Lee.Figure2} such that the crystal orientation can be completely understood prior to the measurement.  Since the effects of charge order or spin order are at most relatively small compared to those associated with intrinsic anisotropy of the material, the absolute orientation of the crystals can be determined from the measured rotation pattern of the full quadrupolar spectrum, obtained using a single-axis goniometer, interpreted through fitting to the exact diagonalization of the nuclear spin Hamiltonian~\cite{Carter1977}.  In this way we were able to align the $c$-axis  close to the external field, $H_0$, with accuracy ($\mathrm{\delta}\theta$,\,$\mathrm{\delta}\phi$) = ($\pm$ 1$^\circ$, $\pm$ 2$^\circ$).  In this range, the expected upper bound for splitting of O(1) into O(1a) and O(1b) is $\sim$\,2 kHz and less than the typical resolution of the NMR experiment, precluding crystal misalignment as the source for any observed frequency shifts.

The frequencies of satellites for less than axial symmetry as for O(1) are given explicitly by a summation of magnetic and quadrupolar contributions in the high magnetic field  limit~\cite{Carter1977},

\begin{equation}
\label{eq1}
f(\theta,\phi,m)=f_{mag}+f_{quad}^{(1)}+f_{quad}^{(2)}
\end{equation}
\begin{equation}
\label{eq2}
f_{mag}=\gamma H_{0} (1+K_{0}+K_{1} \frac{(3\,\mathrm{cos}^2\theta-1)}{2}- K_{2} \frac{\mathrm{sin}^2\theta \,\mathrm{cos}2\phi}{2})
\end{equation}

\begin{equation}
\label{eq3}
f_{quad}^{(1)} =\nu_{Q} (m-\frac{1}{2}) [\frac{(3\,\mathrm{cos}^2\theta-1)}{2} -\eta \frac{\,\mathrm{sin}^2\theta \,\mathrm{cos}2\phi}{2}]
\end{equation}

\noindent
The gyromagnetic ratio for $^{17}$O is $\gamma$ = 5.7719 MHz/T, which we take as the zero reference for the Knight shift, $K_{0}$. The quadrupolar frequency $\it{\nu_{Q}}$  is proportional to the principal component of the EFG,  
\begin{equation}
\label{eq4}
\frac{\partial^2V}{\partial z^2} = \frac{\nu_Q2I(2I-1)h}{3eQ}\geq \frac{\partial^2V}{\partial y^2}\geq \frac{\partial^2V}{\partial x^2}
\end{equation}
\noindent
where $\it{V}$ is the electrostatic potential, Q the quadrupole moment of the nucleus, and  the principal axes, $x$, $y$, and $z$,  for the EFG are chosen to satisfy these inequalities. The three eigenvalues of the EFG tensor are proportional to the quadrupole frequency $\nu_{Q}$ with the $z$-axis along the Cu-O bond direction. The 2$I+1$ quadrupolar perturbed Zeeman substates are indexed by $m$. The superscript in the quadrupolar frequency distinguishes between  1st and 2nd order terms from perturbation theory in the high field limit.  The form of the second order correction,  Supplementary Materials, was included in our full analysis of the spectrum  but has a negligible contribution to the frequency shifts, $\lesssim$\,1 kHz.  The EFG anisotropy parameter is $\eta \equiv (\partial^2V/\partial x^2 - \partial^2V/\partial y^2)/(\partial^2V/\partial z^2)$.   Lastly, the polar and azimuthal angles, $\it{\theta}$ and $\it{\phi}$, denote the orientation of $H_0$ with respect to $x, y, z$-axes. 

For less than axial symmetry, as for O(1), the  components of the Knight shift  are measured with magnetic field along crystal axes,  $K_a$, $K_b$, and $K_c$ and can be expressed as isotropic, $K_0$, and axial  shifts, $K_1$ and $K_2$.  For O(1), $K_{0} = (K_{a}+K_{b}+K_{c})/3;\,\,\,K_{1} = (2K_{a}-K_{b}-K_{c})/3;\,\,\,
K_{2} = K_{b}-K_{c} \equiv \epsilon K_{1}$.  In contrast O(2), the apical oxygen site well-removed from the copper-oxygen plane, has axial symmetry along the $c$-axis and the anisotropy ratio, $\it{\epsilon}$, is zero.  Using $^{17}$O NMR at the apical site of UD74 we reported previously  there was no evidence for static loop currents or for charge or spin ordering~\cite{Mounce2013}, which is also the case for Y123~\cite{Wu2015}.

\begin{figure}[!ht]
\centerline{\includegraphics[width=8.5cm]{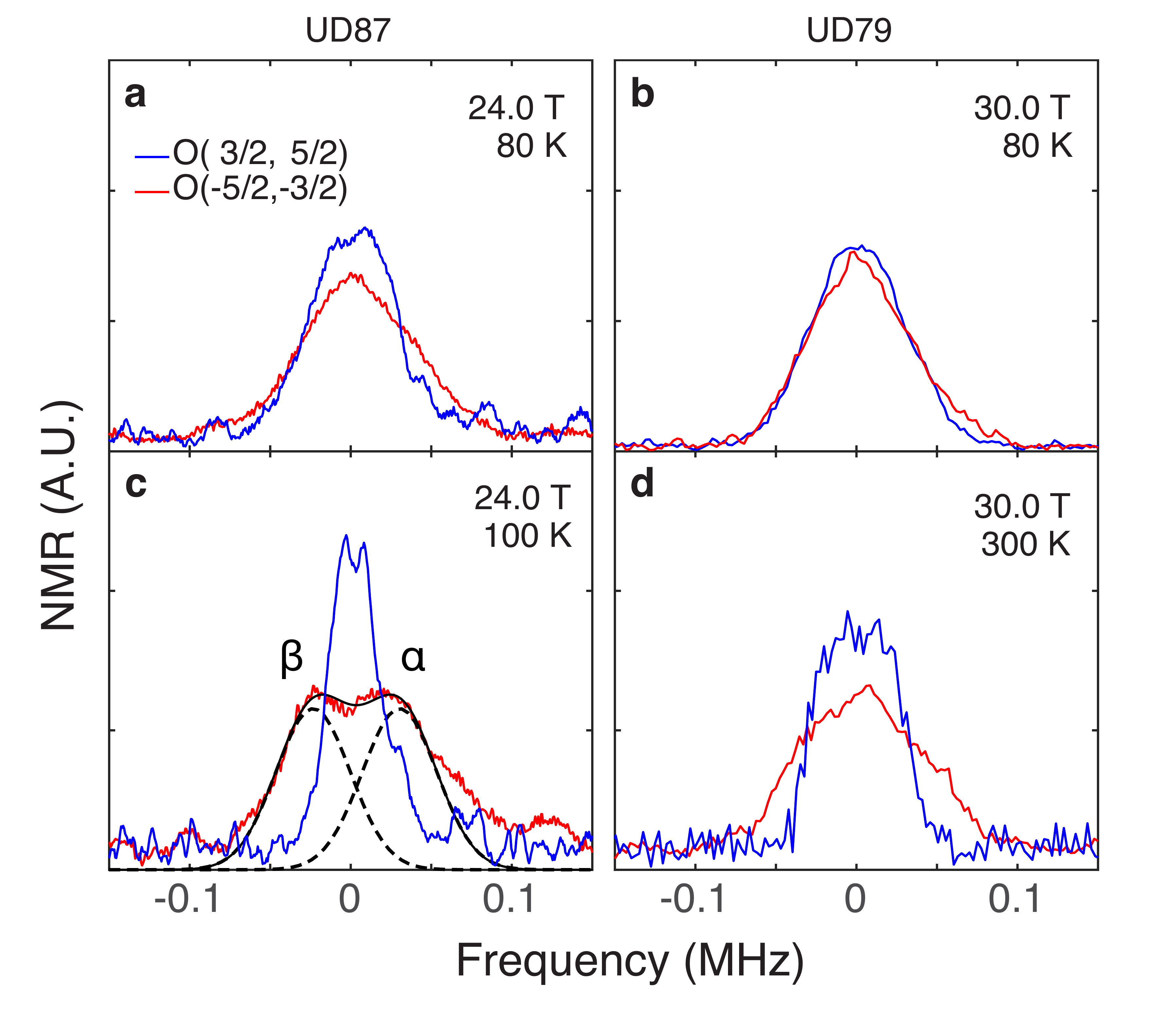}}
\caption {[a, b] Examples of NMR spectra from the highest (red) and the lowest (blue) frequency satellites of the planar oxygen at two temperatures and fields parallel to the $c$-axis. Linewidth asymmetry between upper and lower frequency satellites for O(1), most significant at high temperatures, requires coexistence of spatially resolved spin and charge order. The $\alpha$ peak is defined as the higher frequency component and the $\beta$ peak as the lower one of the doublet, remaining so at all fields and temperatures for all upper satellites.  The same is not true for the lower satellites and the two peaks can merge forming substantially narrower compound spectra, as shown in c).}
\label{Lee.Figure3}
\end{figure}

\begin{figure}[!ht]
\centerline{\includegraphics[width=8.5cm]{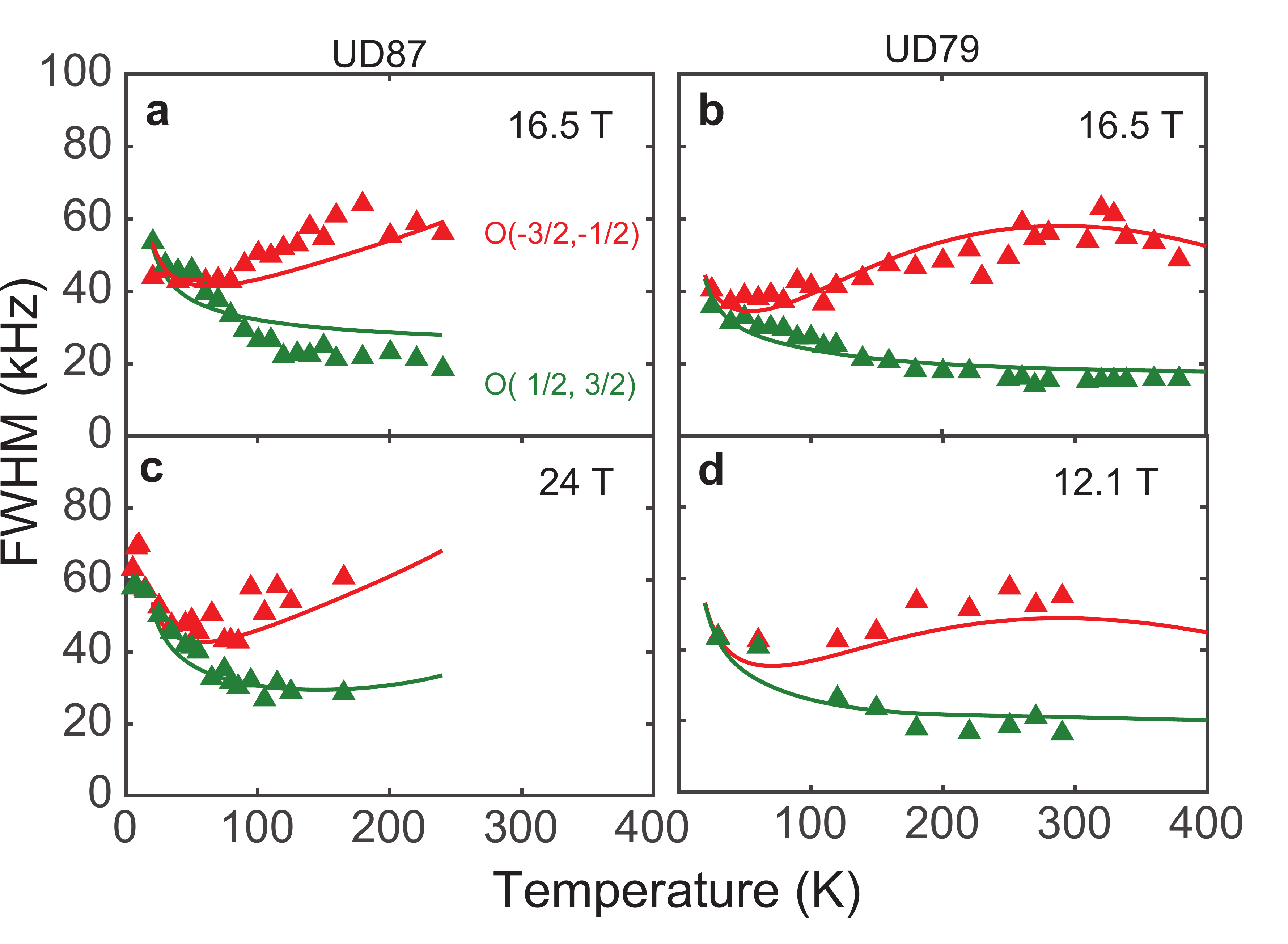}}
\caption {[a, b] Anomalous asymmetry in the O(1)(-1/2,-3/2) and O(1)(3/2,1/2) satellites as a function of temperature expressed by the full-width-at-half-maximum linewidth (FWHM) at various fields from 6.4 T to 30 T. The solid lines are calculated from the parameters obtained from a global fit to the full set of spectral peaks using the model of two inequivalent planar oxygen sites (see text).}
\label{Lee.Figure4}
\end{figure}

However, a close examination of O(1), the planar oxygen spectra, indicates substructure.  In Fig.\ref{Lee.Figure3} there is an unambiguous  splitting of the highest frequency satellite in our  UD87 sample  at temperatures above 100\,K, in contrast to the corresponding lowest frequency satellite which has a much narrower linewidth.   The spectrum at $T = 180$\,K can be separated into two gaussian shaped peaks of equal weights ($\alpha$ and $\beta$). Consistent behavior was observed in  UD79.  Above 100\,K we  found that the full-width-at-half-maxima (FWHM) of the upper satellites were  greater than for the lower satellites, Fig.~\ref{Lee.Figure4}.  We take this as  evidence for such a splitting even though it is unresolved.   The asymmetry in the linewidths, {\it i.e.} comparison of high and low frequency satellites, can  be accounted for by simultaneous spin and charge spatial modulations.  Electronic order in only one of spin or charge channels is insufficient.  Similar observations were reported by Wu \et\ ~\cite{Wu2011,Wu2013} from both $^{63}$Cu and $^{17}$O NMR for chain ordered Y123, and was interpreted as evidence for charge ordering, magnetic field induced ordering in that case.  The maximum $^{17}$O spectral splitting  we have found in Hg1201 is $\sim$\,60\,kHz, similar to that reported for Y123 ~\cite{Wu2013} where $\nu_Q \sim$\,0.9\,MHz.

Asymmetric distributions of quadrupolar satellite lineshapes can only arise  under rather restrictive  circumstances apart from possible electronic ordering. For example, misalignment of crystals relative to the magnetic field, or the existence of bi-crystal domains can result in an NMR spectral splitting.  We have looked at these possibilities using different models for analysis, but find that they do not account for our spectra (see Supplementary Materials).

Consequently, it is most likely that the  inequivalency of  oxygen sites that we observe has a local origin  associated with coexisting charge and spin density waves.  The fact that the NMR satellites are asymmetrically affected requires that these modulations coexist, that they are coherent, and finally that they are locked in phase, as demonstrated by simulations in Supplementary Materials.  Independent or out of phase superpositions of spin and charge order averaged over the sample destroy the asymmetry and broaden the NMR spectrum. Being intimately linked they must have the same origin.  For simplicity we have analyzed our spectra in the context of a simple model of two inequivalent oxygen sites $\alpha$ and $\beta$ in the CuO$_2$ plane having equal weight.  Even if a spin or charge periodic order is incommensurate with the lattice it will be manifest in a prominent two-peak spectrum.  In the following we describe our results using our model to determine the quadrupole frequency  and Knight shift splittings which are a measure of the charge  and spin density wave amplitudes respectively. 

\begin{figure}[!ht]
\centerline{\includegraphics[width=8.5cm]{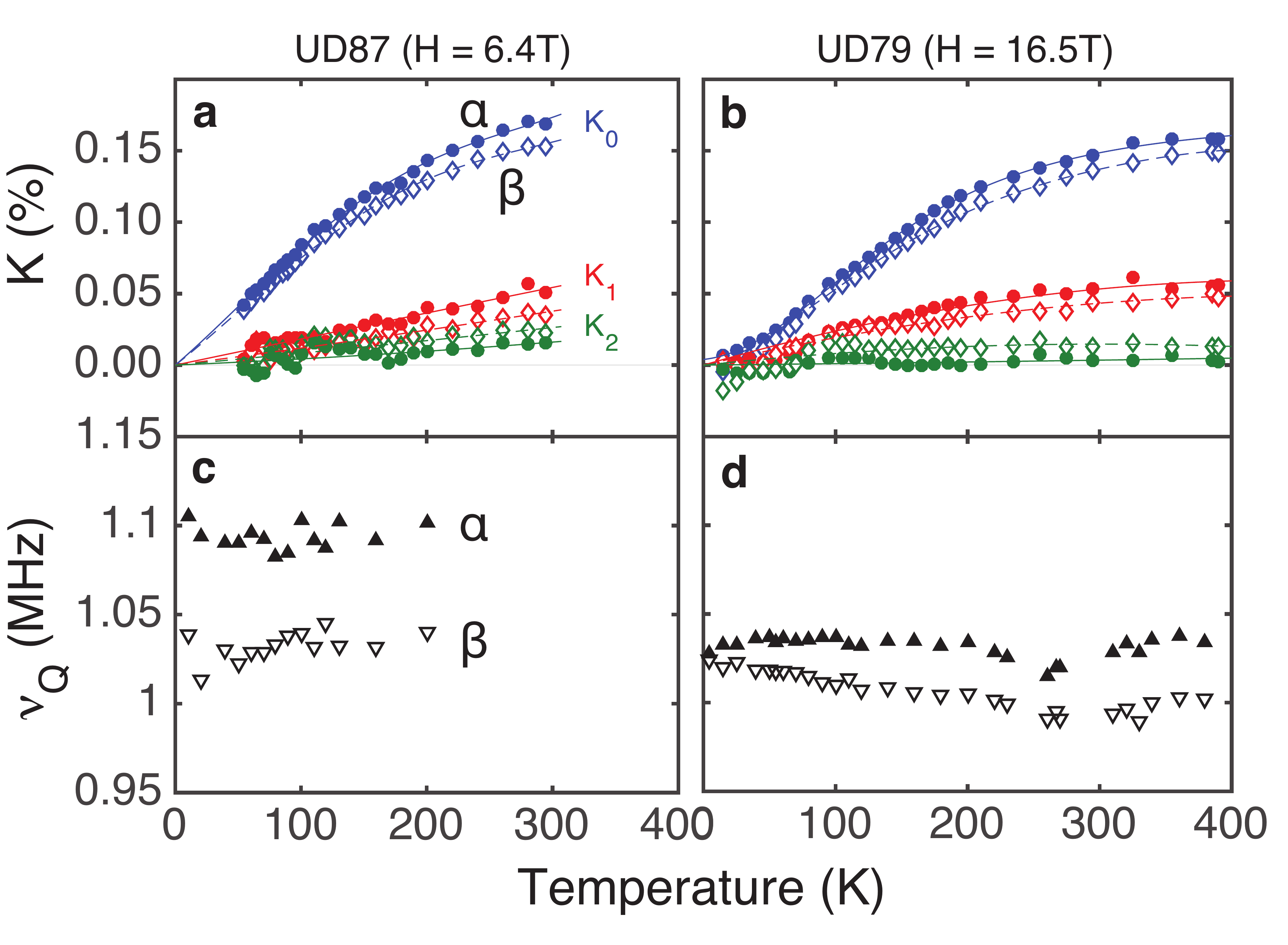}}
\caption {[a, b] Temperature dependence of Knight shifts of the two inequivalent sites. The lines are guides to the eye. Error bars are smaller than the separation between $\alpha$ and $\beta$ data at most temperatures. [c, d]  The separation of the satellites, $\nu_{Q}$,  is plotted as a function of temperature. }
\label{Lee.Figure5}
\end{figure}

The uncertainties in determining frequency splittings from unresolved overlapping peaks is reduced by imposing mathematical constraints on the ten-gaussian fitting.  We assumed that $\alpha$ and $\beta$ peaks are gaussians of equal weight and width and that the magnetic and quadrupolar contributions are added in quadrature. We observed  that the  average Knight shift position of $\alpha$ and $\beta$ peaks was magnetic field independent at two fields, $H_0 =16.5$ and 30\,T.  Consequently, we constrained each Knight shift component to be magnetic field independent.  A discussion of this constraint and the consideration of alternate models is presented in Supplementary Materials.  Analysis of all spectra was performed by fitting as required by Eq.\ref{eq1}-\ref{eq3}.  The NMR fit parameters are: $\nu_{Q,\alpha} , \nu_{Q,\beta}, K_{0,\alpha}, K_{1,\alpha}, K_{2,\alpha},K_{0,\beta}, K_{1,\beta}, K_{2,\beta}, \sigma_{m},\sigma_{q}$ where the last two are magnetic and quadrupolar linewidths.  

The results for several cases are shown in Fig.~\ref{Lee.Figure5} and in Supplementary Materials for $\sigma_q$ and  $\sigma_m$.  We found that the Knight shifts were only weakly doping dependent, that $\sigma_{q}$ was temperature independent  and independent of field from 6.4 to 30 T, and  that $\sigma_{m}/H_0$ followed a Curie-Weiss law. Using the fitting parameters,  we were able to accurately simulate the spectra in Fig.~\ref{Lee.Figure2}.  With these parameters, we calculated the FWHM linewidth  of each composite peak shown as solid curves  in Fig.~\ref{Lee.Figure4}. The consistency with the raw data confirms the appropriateness for our model of two inequivalent oxygen sites.  In addition, the partial splitting  at the highest frequency satellite by $\sim$\,60\,kHz at $H_0 = 30$\,T, shown in Fig.~\ref{Lee.Figure3}\,c), agrees quantitatively with the calculated spectra based on the fitted parameters given by the solid black curve.  
\begin{figure}[!ht]
\centerline{\includegraphics[width=8.5cm]{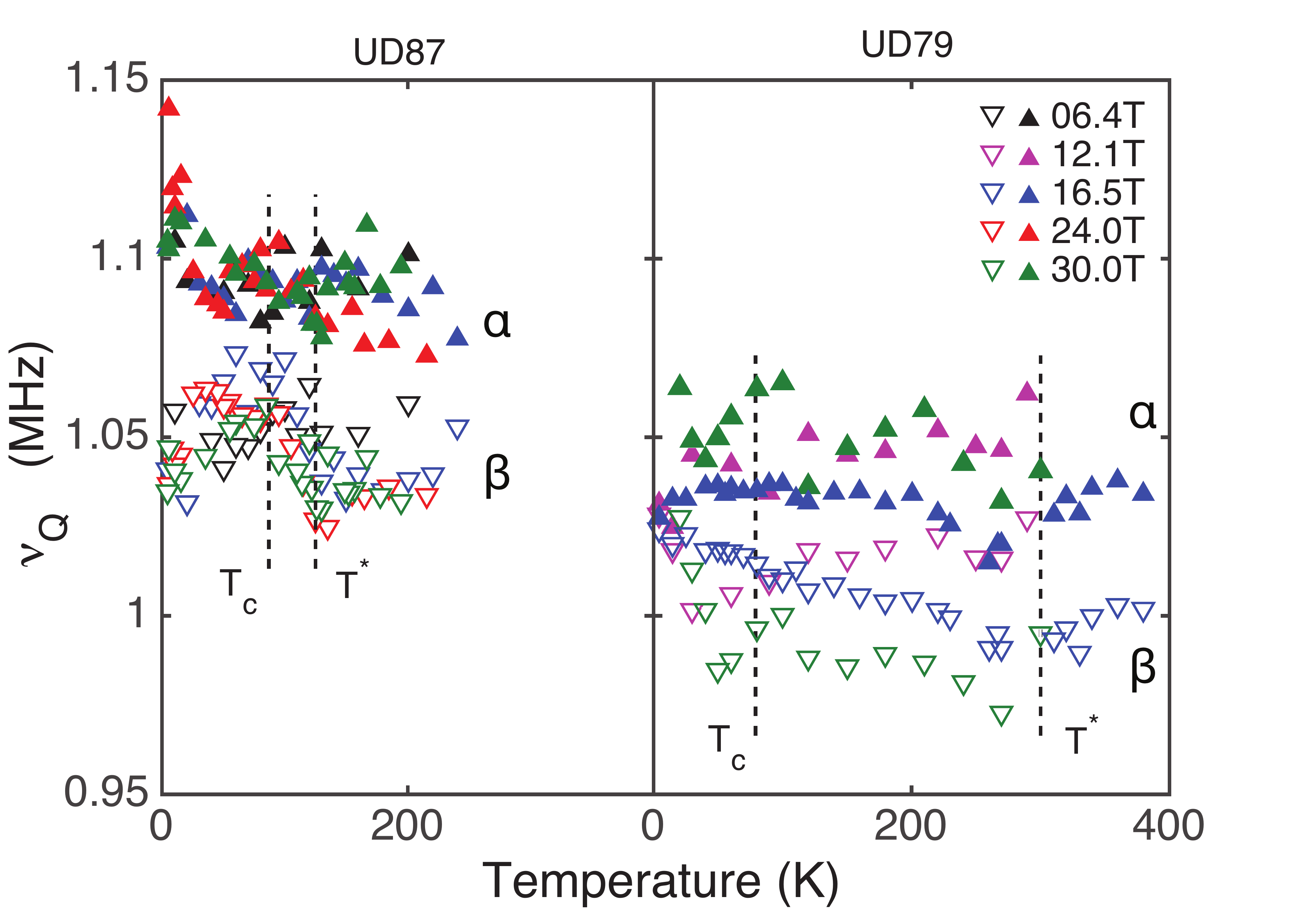}}
\caption {Temperature and field dependence of $\nu_Q$ is consistent with the existence of charge order manifest in the difference between $\alpha$ and $\beta$ sites.  But there is no evidence for a field or temperature induced onset. Statistical errors in the plot are about $\sim13$\,kHz except the H = 16.5\,T data for  UD79 which has $\sim$\,3 kHz where longer measurement times were possible.  For a third sample, UD74, we only have data at 100 K but it is consistent with the above with $\Delta\nu_Q = 40$\,kHz at both 21 and 24\,T.  The transition temperatures and pseudogap temperatures are shown by dashed vertical lines, T$^{*}$ = 125\,K for UD79, and  300\,K for UD87, determined from our $^{63}$Cu NMR T$_{1}$ relaxation measurements~\cite{Lee2016}.}
\label{Lee.Figure6}
\end{figure}

The quadrupolar frequency of each site is shown in Fig.~\ref{Lee.Figure6} for all our data.  A uniform field-independent difference $\Delta \nu_Q \equiv \nu_{Q,\alpha} - \nu_{Q,\beta} \sim$\,40 kHz for $\alpha$ and $\beta$ sites above $\sim$\,100\,K, is evident.  In this range our measurements in Hg1201 indicate static electronic order. The general trend of increasing $\nu_{Q} $ with decreasing temperature can be attributed to temperature dependence of the lattice parameters.  This interpretation can be made with more confidence using the apical site data; see Supplementary Materials. The magnetic and quadrupolar frequency shifts of $^{17}$O NMR in cuprates  can be associated with an intrinsic electronic instability with coherent charge and spin order in the CuO$_2$ plane. However, there is no evidence for a field or temperature induced onset, nor is there any  correlation with vortex freezing (Fig.~\ref{Lee.Figure1}) or with the pseudogap.  The decrease in $\nu_Q$ in UD87 below T$_{c}$ might be a consequence of suppression of the CDW indicated by x-ray measurements \cite{Chang2012,Ghiringhelli2012,Comin2015}; however, this effect is not evident in sample UD79. It was suggested \cite{Wu2013} that electronic ordering in Y123  fluctuates and that it is observed by NMR only when pinned, leading to an onset induced by a combination of sufficiently high magnetic field and sufficiently low temperature possibly associated with vortex freezing.  This explanation does not account for magnetic field induced charge order observed in x-ray diffraction~\cite{Gerber2015}, nor does it correspond to our data for Hg1201.  Identification of an onset of charge order, with no complementary spin order in Hg1201 was reported from resonant x-ray scattering~\cite{Tabis2014} at the copper L$_3$-edge. This comparison provides a complementary view of charge ordering  at different positions in the same material probed on different time scales. From x-ray diffraction in Hg1201 at optimal doping~\cite{Campi2015} evidence was found that oxygen interstitials induce  formation of charge puddles in the CuO$_2$ plane, observations possibly related to the oxygen site inequivalency we report here.

{\bf Acknowledgements.}  We acknowledge contributions from Andrew Mounce, discussions with Marc-Henri Julien, and Vesna Mitrovic,  and we thank Martin Greven for supplying the unprocessed Hg1201 crystals.   Research was supported by the U.S. Department of Energy, Office of Basic Energy Sciences, Division of Materials Sciences and Engineering under Awards DE-FG02-05ER46248 (Northwestern University).  M.K.C. is supported by funds from the US Department of Energy BES grant no. LANLF100.  A portion of this work was performed at the National High Magnetic Field Laboratory, which is supported by National Science Foundation Cooperative Agreement No. DMR-1157490 and the State of Florida.

\bibliographystyle{naturemag}
\bibliography{ChargeOrder}

\end{document}


\title{Supplementary Materials:  Coherent Charge and Spin Density Waves in Underdoped HgBa$_{2}$CuO$_{4+\delta}$}

\author{Jeongseop A. Lee$^{1}$, Yizhou Xin$^{1}$, W. P. Halperin$^{1}$, A. P. Reyes$^{2}$ P. L. Kuhns$^{2}$, M. K. Chan$^{3}$}

\affiliation{$^1$Department of Physics and Astronomy Northwestern University, Evanston, Illinois 60208, USA\\$^2$National High Magnetic Field Laboratory, Tallahassee, Florida 32310, USA\\$^3$Pulsed Field Facility, National High Magnetic Field Laboratory, Los Alamos National Laboratory, Los Alamos, New Mexico 87545, USA}

\baselineskip24pt

\maketitle 
\newcommand{\beginsupplement}{%
        \setcounter{table}{0}
        \renewcommand{\thetable}{S\arabic{table}}%
        \setcounter{figure}{0}
        \renewcommand{\thefigure}{S\arabic{figure}}%
        \setcounter{equation}{0}
        \renewcommand{\theequation}{S\arabic{equation}}%
     }
\beginsupplement

\section{I. Quadrupolar Interaction}
In the high field limit the quadrupolar interaction can be expanded for each eigenstate of the Zeeman Hamiltonian using perturbation theory. The first order quadrupolar interaction for transition (m, m-1) is given in  Eq. (3). The second order interaction is described by the following equation.

\begin{eqnarray}
\begin{aligned}
\label{eqS1}
f_{quad}^{(2)} &=  \frac{\nu_{Q}^{2}}{H_{0}\gamma} (1-\mathrm{cos}^{2}\theta)\times  [\{102m(m-1)-18I(I+1)+39\} \mathrm{cos}^{2}\theta (1+\frac{2}{3}\eta\,\mathrm{cos}2\phi) \\
&-\{6m(m-1)-2I(I+1)+3\} (1-\frac{2}{3}\eta\,\mathrm{cos}2\phi)] \\
&+\frac{\eta^{2}}{72}\frac{\nu_{Q}^{2}}{H_{0}\gamma}[24m(m-1)-4I(I+1)+9 \\
& -\{30m(m-1)-2I(I+1)+12\}\mathrm{cos}^{2}\theta \\
&-\{\frac{51}{2}m(m-1)-\frac{9}{4}I(I+1)+\frac{39}{4}\} \mathrm{cos}^22\phi(\mathrm{cos}^2\theta-1)^2].
\end{aligned}
\end{eqnarray}
\\
\noindent
The magnitude of the second order term varies with the transition orders but it is generally less than $\sim$ 1 kHz for the sample orientation, $\it{H_0}$ $\parallel$ c. Our analysis in the main text, as well as our calculation of the linewidths shown in Fig. 4, include this second order interaction term.  The angular dependence of all  NMR peak frequencies is consistent with the known tetragonal structure of Hg1201 (Fig. 2), and the crystal parameters extracted from this angular dependence are given in table \ref{tableS1}.

\begin{table}[!th]
\centering
\begin{tabular}{|c|c|c|c|}
\hline
Sample & UD74K &  UD79 &  UD87 \\
\hline
\hline
K$_{0}$ (\%)& 0.056 & 0.055 & 0.078  \\
\hline
K$_{1}$ (\%)& 0.018 & 0.0214 & 0.015  \\
\hline
K$_{2}$ (\%)& -0.01 & 0.0045 & 0.006  \\
\hline
$\nu_{Q}$ (MHz)& 1.05 & 1.0174 & 1.05  \\
\hline
$\eta$ & 0.386 & 0.3738 & 0.385  \\
\hline
\end{tabular}
\caption{Knight shift and quadrupolar frequencies for the planar oxygen in underdoped Hg1201 single crystals at H = 16.5\,T and T = 80\,K (UD79) and T = 100\,K (UD87). These results are associated with the average peak positions of $\alpha$ and $\beta$ sites. Results for the underdoped T$_{c}$ = 74\,K (UD74K) crystal have been reported previously~\cite{Mounce2013}, including parameters for the apical oxygen (green solid curves in Fig. 2).}
\label{tableS1}
\end{table}

\section{II. Two-Site Model}
In order to explain the asymmetric linewidth distribution between the quadrupolar satellites, we used a two-site model  assuming two almost degenerate oxygen sites, each characterized by a unique set of magnetic and quadrupolar parameters:
$\nu_{Q,\alpha}$, $\nu_{Q,\beta}$, $K_{0,\alpha}$, $K_{1,\alpha}$, $K_{2,\alpha}$, $K_{0,\beta}$, $K_{1,\beta}$, $K_{2,\beta}$, $\sigma_{m}$, $\sigma_{q}$. The two sites are indicated by subscripts $\alpha$ and $\beta$.  We assume that the principal axes for the proposed sites are identical since we have independently investigated this question with a second approach discussed next in section III called the bi-crystal tilt model, and found that it could not account for our spectra.  We argued in the main text that the magnetic  frequency shifts could be described by magnetic field independent Knight shifts. This will be explored further in section IV where we conclude that an alternative model of a field independent spin density wave cannot be excluded. 

\begin{figure}[!ht]
\centerline{\includegraphics[width=10cm]{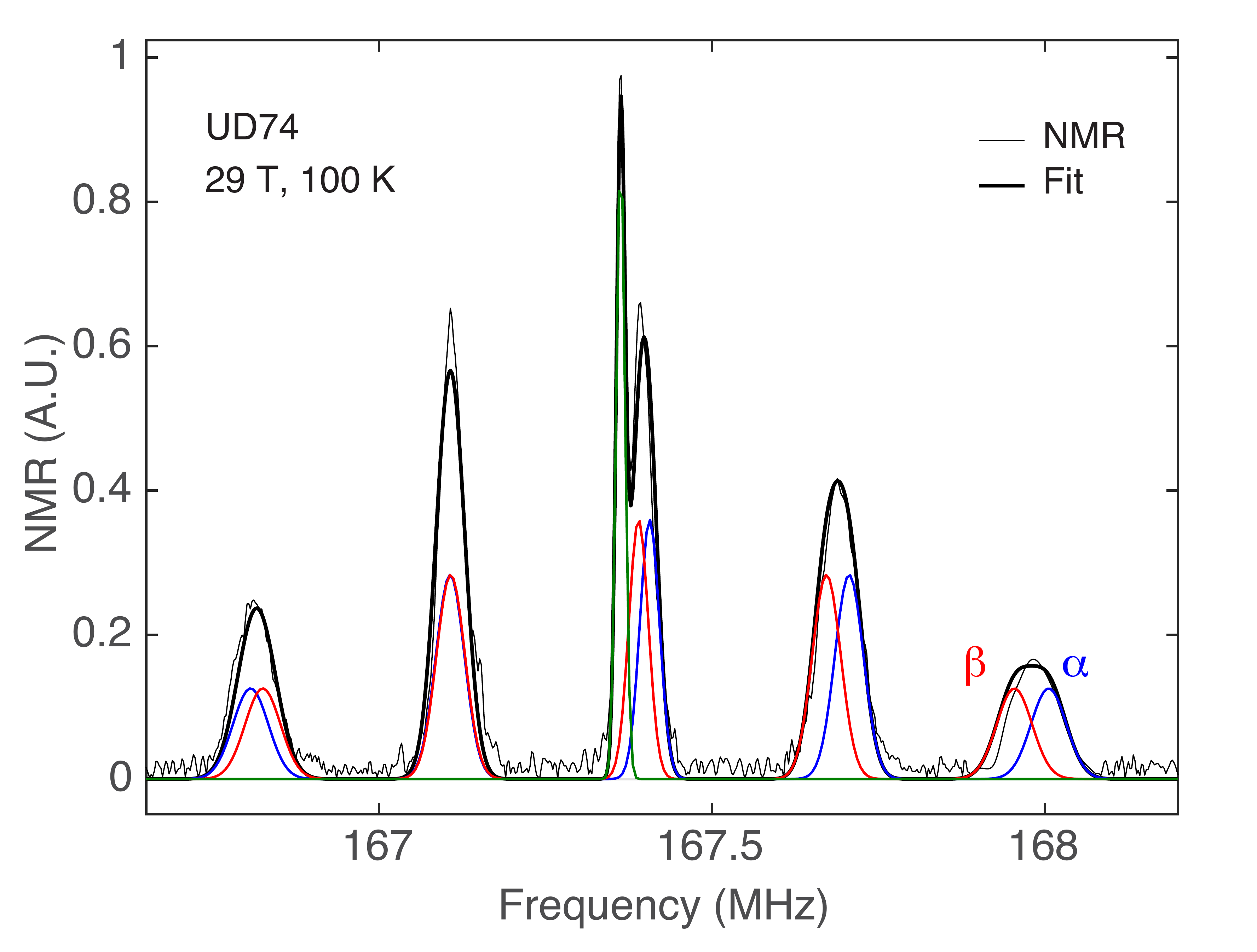}}
\caption {Two-site model fitting  for UD74 spectra at 29\,T and 100\,K and $H_{0}||c$. The observed linewidth asymmetry is generated by a slight mismatch in magnetic and quadrupolar parameters between the two sites.  The asymmetry is  easily identified from the variation in the maximum intensity of the quadrupolar peaks. A sharp peak in the central site (green) is from the apical oxygen whose quadrupolar satellites occur with a larger $\nu_{Q}$ than can be displayed here.}
\label{Lee.FigureS1}
\end{figure}

\begin{equation}
\begin{split}
\label{eqS2}
A(f) = & \sum_{m,i} A_{(m,m-1)} exp\bigg(-\frac{(f-(f_{mag,i}(\theta,\phi)+f_{quad,i}(m,\theta,\phi)))^2}{2(\sigma_m^2+(\lvert m-1/2\rvert \sigma_q)^2)}\bigg).  \end{split}
\end{equation}

Eq.~\ref{eqS2} represents  NMR spectra  for $I\,=\,\frac{5}{2}$ consistent with the quadrupolar Hamiltonian to first order  for two inequivalent sites $i=\alpha,\,\beta$. The NMR amplitudes for symmetric pairs of upper and lower quadrupolar satellites are set to be equal, i.e., $\it{A}$$_{(-3/2,-1/2)}$ $\it{=}$ $\it{A}$$_{(1/2,3/2)}$ and $\it{A}$$_{(-5/2,-3/2)}$ $\it{=}$ $\it{A}$$_{(3/2,5/2)}$, as is appropriate for a field sweep NMR experiment. The centroid of a quadrupolar pattern and the quadrupolar separations are given by, $f_{mag,i}$ and $f_{quad,i}$, with additional subscript designation for site distinction since each site generates its own full quadrupolar pattern. The magnetic and quadrupolar contributions to the total linewidth are represented as $\sigma_m$ and $\sigma_q$, respectively, where each spectral peak is a gaussian with width $\sigma$ taken as the sum in quadrature of the two contributions and where we note that the quadrupolar term is linearly incremented as the transition order increases, providing an unambiguous distinction between these two contributions. To calculate the FWHM linewidth we use 2.36$\, \it{\sigma}$.

\section{III. Bi-crystal Tilt Model}

\begin{figure}[!ht]
\centerline{\includegraphics[width=10cm]{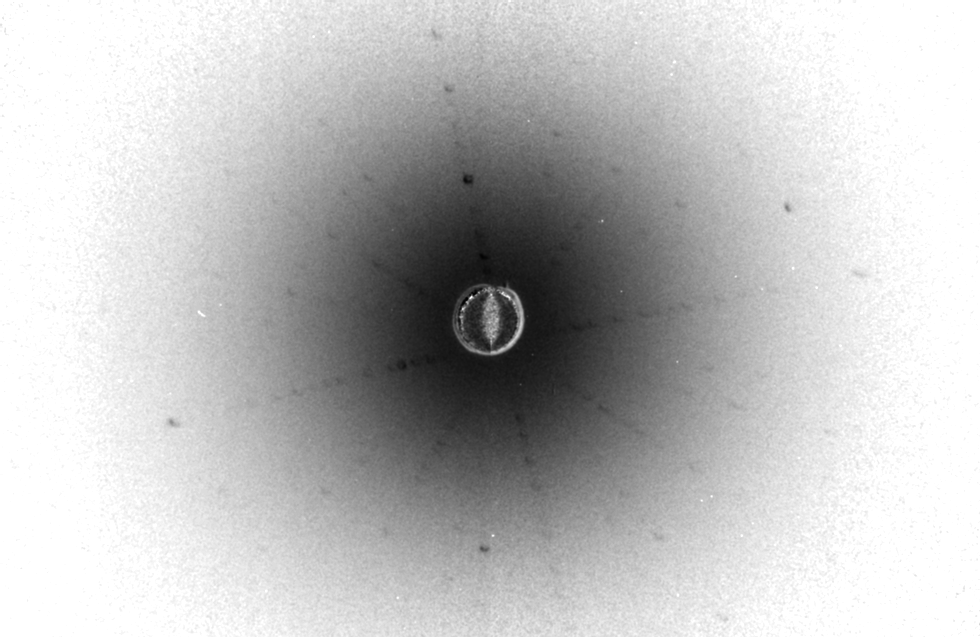}}
\caption {Laue image of the UD87 sample normal to the $c$-axis shows a single crystal domain consistent with the crystal structure of Hg1201. We have found no evidence of multiple crystal domains in the cation framework in any of our samples.  However, the x-ray Laue method is not sensitive to oxygen.}
\label{Lee.FigureS2}
\end{figure}

The less than axial symmetry of the planar oxygen in Hg1201 can result in multiple quadrupolar patterns if the external field is not precisely along the c-axis since O(1a) and O(1b) are no longer degenerate under those conditions. The most straightforward manifestation of this phenomenon in a single crystal is a simple misalignment of the magnetic field orientation. We can eliminate this possibility using a goniometer to map a rotation pattern and align the crystal {\it in-situ} with high precision. If the sample consists of a bi-crystal, {\it i.e.} two crystal domains, this could also produce a splitting of the spectra.  But the Laue diffraction images of our single crystals exclude existence of multiple crystal domains, Fig. \ref{Lee.FigureS2}. 

Other mechanisms include existence of isotope-induced buckling of the Cu-O plaquette.  There have been reports of isotope-induced buckling modes of the octahedra~\cite{Kirk1988,Crawford1990,Pickett1991,Jepsen1998,Chmaissem1999,Dahm2000,Bussmann-Holder2005,Iwasawa2008}. Since we have isotope-exchanged our samples, we explored this possibility by attempting to fit the spectra with a ten-gaussian expression similar to the one in our two site model. Instead of allowing two distinct sets of magnetic and quadrupolar parameters, we  constrained the fitting to a single set of parameters for both sites while allowing for a dictinction between O(1a) and O(1b) to be manifested through different relative orientations of the principal axes for each crystal domain. The key difference between the two-site model and the bi-crystal tilt model discussed in this section is that in the latter there are two distinct sets of principal axes whose relative orientations with respect to the magnetic field direction are expressed in terms of the spherical coordinates, ($\theta_{i},\phi_{i}$) as fit parameters which are different for $i=1,2$ where the index $i$ specifies one of the two principal axis domains.

This model does not fit our experimental spectra. In general, the compound spectra of a nearly degenerate pair of sites with the same principal axes (two-site model) has a larger linewidth for upper quadrupolar satellites only if the site with a larger magnetic shift also exhibits a larger quadrupolar shift so that both shift components constructively contribute to the total linewidth of the upper quadrupolar satellites. This is in contrast to the lower frequency satellites in which they compensate for one another resulting in a narrower linewidth. The bi-crystal tilt model mostly predicts that a site with a larger magnetic shift has a smaller quadrupolar shift. This results in larger linewidth for lower quadrupolar satellites inconsistent with our experiments. Additionally, a DFT simulation for Hg1201 excludes the existence of octahedra buckling modes which could otherwise generate a relevant scenario appropriate for the bi-crystal tilt model~\cite{Gu2015}.

\section{IV. Simulation of Coherent Incommensurate Density Waves}

The effect of incommensurate spin and charge modulations in the Cu-O plane can be studied from simulation of NMR spectra. We consider 1D spin and charge density waves in terms of the following equations.

\begin{equation}
\label{eqS3}
A_{SDW}(x) = A_{m} \,\mathrm{sin}(2\pi x/\lambda_{m}+\phi_m)
\end{equation}
\begin{equation}
\label{eqS4}
A_{CDW}(x) = A_{q} \,\mathrm{sin}\,(2\pi x/\lambda_{q}+\phi_q)
\end{equation}

The amplitudes of spin and charge density waves (SDW and CDW) are denoted by $\it{A_m}$ and $\it{A_q}$ where $\it{x}$ is the distance in real space. For a coherent density wave modulation, $\it{\lambda_{m}}$ = $\it{\lambda_{q}}$ = $\it{\lambda}$. Incommensurability further requires $\it{\lambda}$ to be an irrational multiple of the lattice constant. The phase of each wave is denoted by $\phi_m$ and $\phi_q$ giving a relative phase of $\phi = \phi_m - \phi_q $.

Density wave modulations induce a doublet splitting for all NMR peaks, corresponding to the $\alpha$ and $\beta$ sites in the two-site model. Our numerical simulation of the effects of the spin and charge density waves in real space was sufficiently large that all the nuclei in a simulation cell provide ample sampling of the  full spatial variation of the density waves. In the case of incommensurate modulations we define the probability density for frequency shifts that a nucleus experiences as follows:

\begin{equation}
\label{eqS5}
P_{SDW}(H_{SDW}) = \int_{}^{} \delta(H_{SDW} - SDW(x)) dx
\end{equation}
\begin{equation}
\label{eqS6}
P_{CDW}(\nu_{Q,CDW}) = \int_{}^{} \delta(\nu_{Q,CDW} - \kappa \frac{\partial}{\partial x}{CDW(x)}) dx.
\end{equation}
where the crystal EFG  scales linearly with $\it{\nu_{Q}}$ in the following equation,
\begin{equation}
\label{eqS7}
\kappa = \frac{EFG}{\nu_Q} = \frac{2I(2I-1)h}{3eQ}.
\end{equation}
\noindent
A precise value of $\lambda$ is not important as long as it is incommensurate with the lattice. A full quadrupolar NMR spectrum was simulated by calculating the spectrum using Eq. (1) but with the frequency of the magnetic and quadrupolar terms perturbed by $\gamma \it{A_{SDW}}$ and $\it{\nu_{Q,CDW}}$, with probability distributions given by Eq. \ref{eqS5} and Eq. \ref{eqS6}, respectively.

\begin{figure}[!ht]
\centerline{\includegraphics[width=10cm]{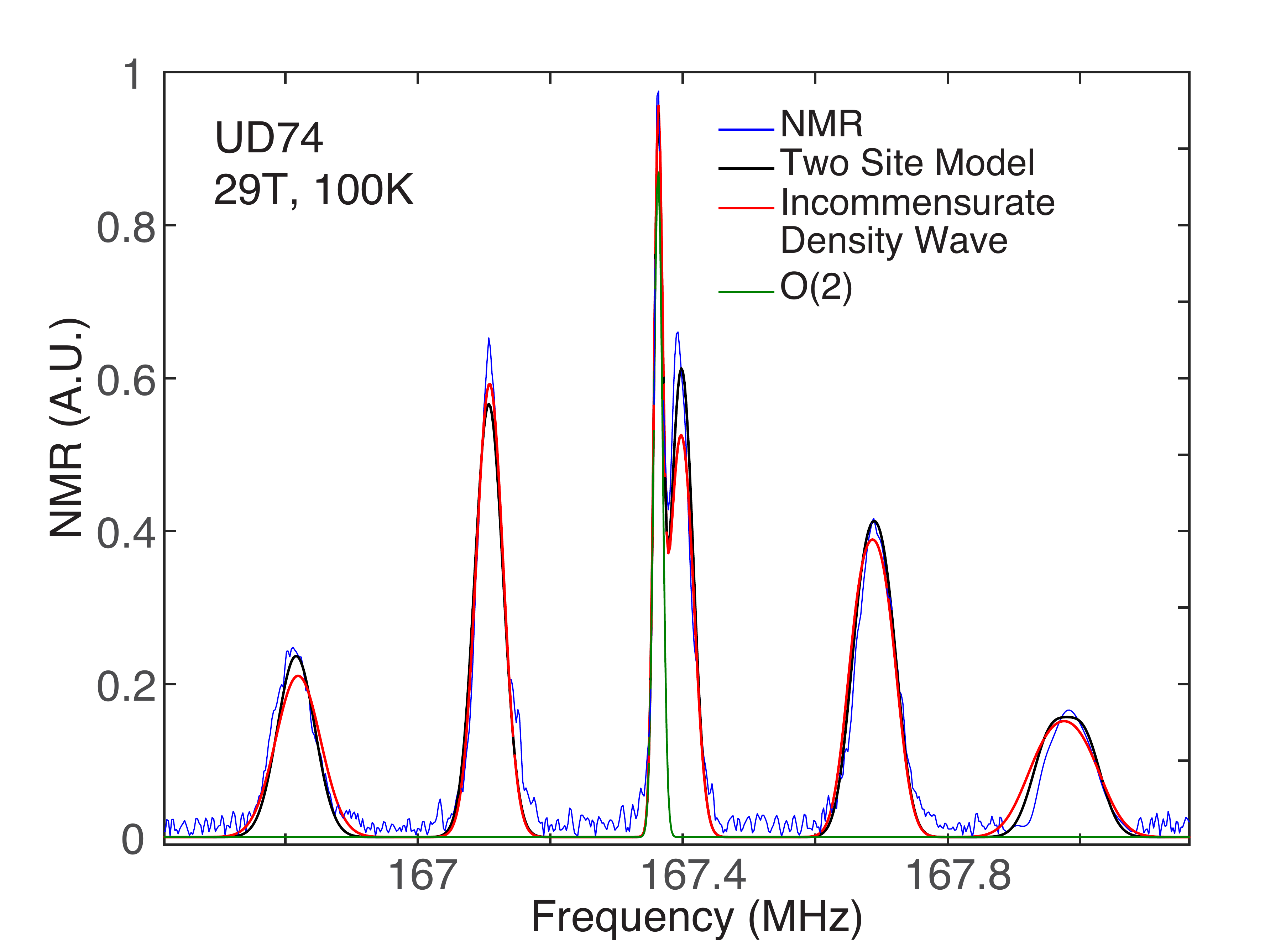}}
\caption {Comparison of the two-site model (black) introduced in the main text and the incommensurate density wave simulation model (red). Both models can be fit to the experimental spectra with similar precision and the asymmetric lineshape distribution among the satellites is accurately represented.}
\label{Lee.FigureS3}
\end{figure}

\noindent
We have fit all our data using this fitting procedure. The fitting variables are, $K_{0}$, $K_{1}$, $K_{2}$, $\nu_Q$, $\sigma_m$, $\sigma_q$, $A_m$, $A_q$, and $\phi$, where the parameter, $\lambda$, is incommensurate with the lattice. 

\begin{figure}[!ht]
\centerline{\includegraphics[width=10cm]{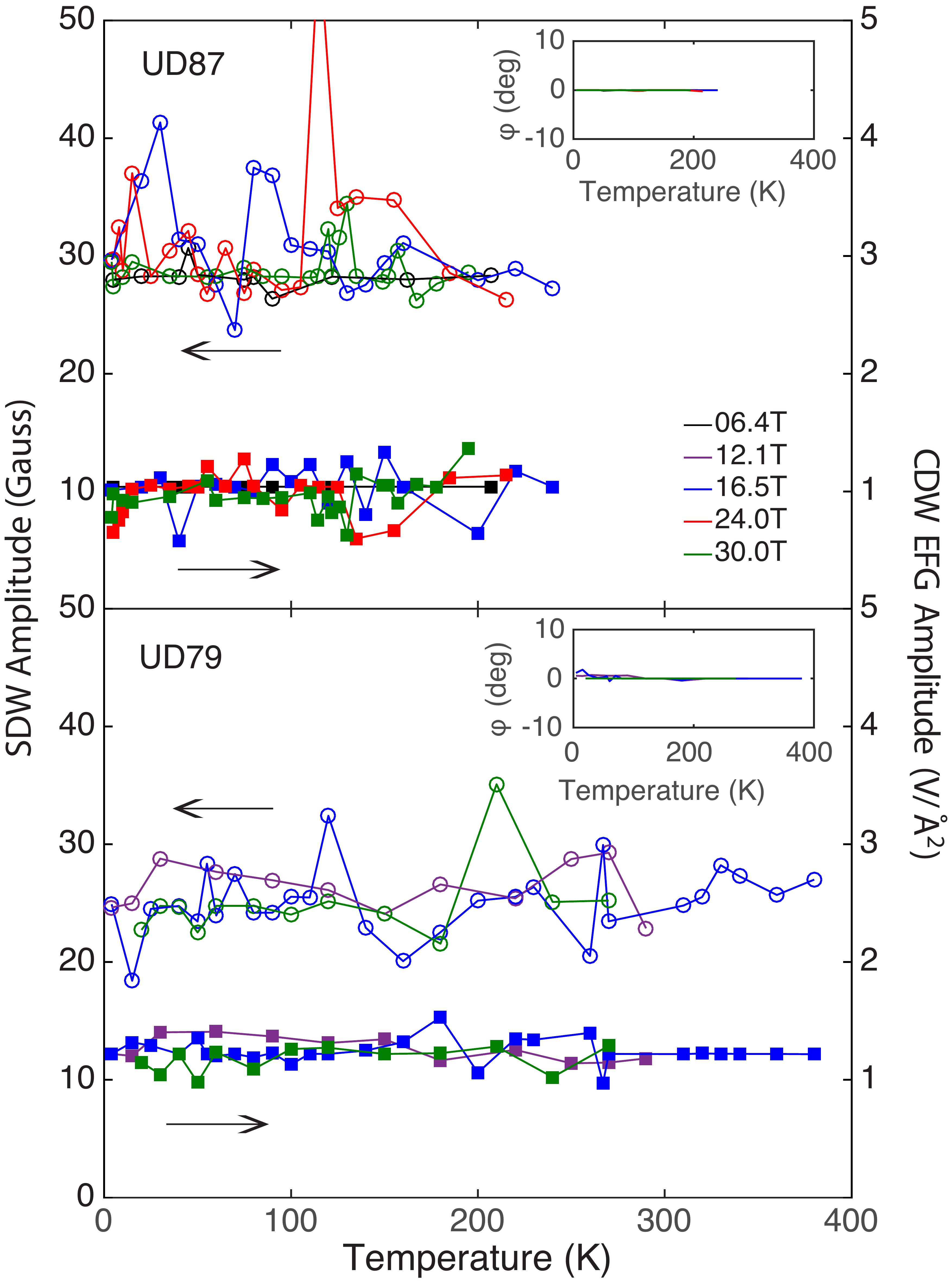}}
\caption{Amplitude of the SDW, $A_m$, and Amplitude of the EFG, $2\pi A_q/\lambda$, determined from fitting all of the experimental spectra using the coherent incommensurate density wave model. This analysis show that both modulations are magnetic field and temperature independent with a well defined phase difference, $\it{\phi}$, which is vanishingly small, indicating that CDW and SDW have a $\it{\pi}$/2 phase difference.}
\label{Lee.FigureS4}
\end{figure}

This model gives new insights. The relative phase angle, $\phi$, and the actual magnitudes of SDW and CDW, $A_m$ and $A_q$, can be determined, although interestingly, they have not been discussed in previous work. According to our analysis, the relative phase $\pi$/2, is very robustly defined; see insets to Fig.~\ref{Lee.FigureS4}.  This phase mismatch between electronic density and spin density modulations has been predicted and observed in rather different circumstances, where it is related to Friedel oscillations \cite{Jena1978,Manninen1981,Yu1981}. The magnitude of the EFG oscillation was found to be about 1 V/$\AA$,  less than 1 $\%$ of the static EFG at the planar oxygen site in Hg1201. The magnitude of the SDW oscillation is  $\sim27$ Gauss, about an order of magnitude less than the incommensurate SDW observed previously in iron-pnictide superconductors\cite{Oh2013}, and it might be related to the magnetism detected by polarized neutron scattering in the same material~\cite{Li2008}.

The accuracies of our fitting analysis to the experimental spectra for the two-site model and the incommensurate density wave model have similar statistics, {\it i.e.}, $\chi^{2}$, as shown in Fig. \ref{Lee.FigureS3}. Therefore, unfortunately, we cannot discriminate between them based solely on the fitting statistics. 

\section{V. Linewidths of $\alpha$ and $\beta$ Components}
\begin{figure}[!ht]
\centerline{\includegraphics[width=10cm]{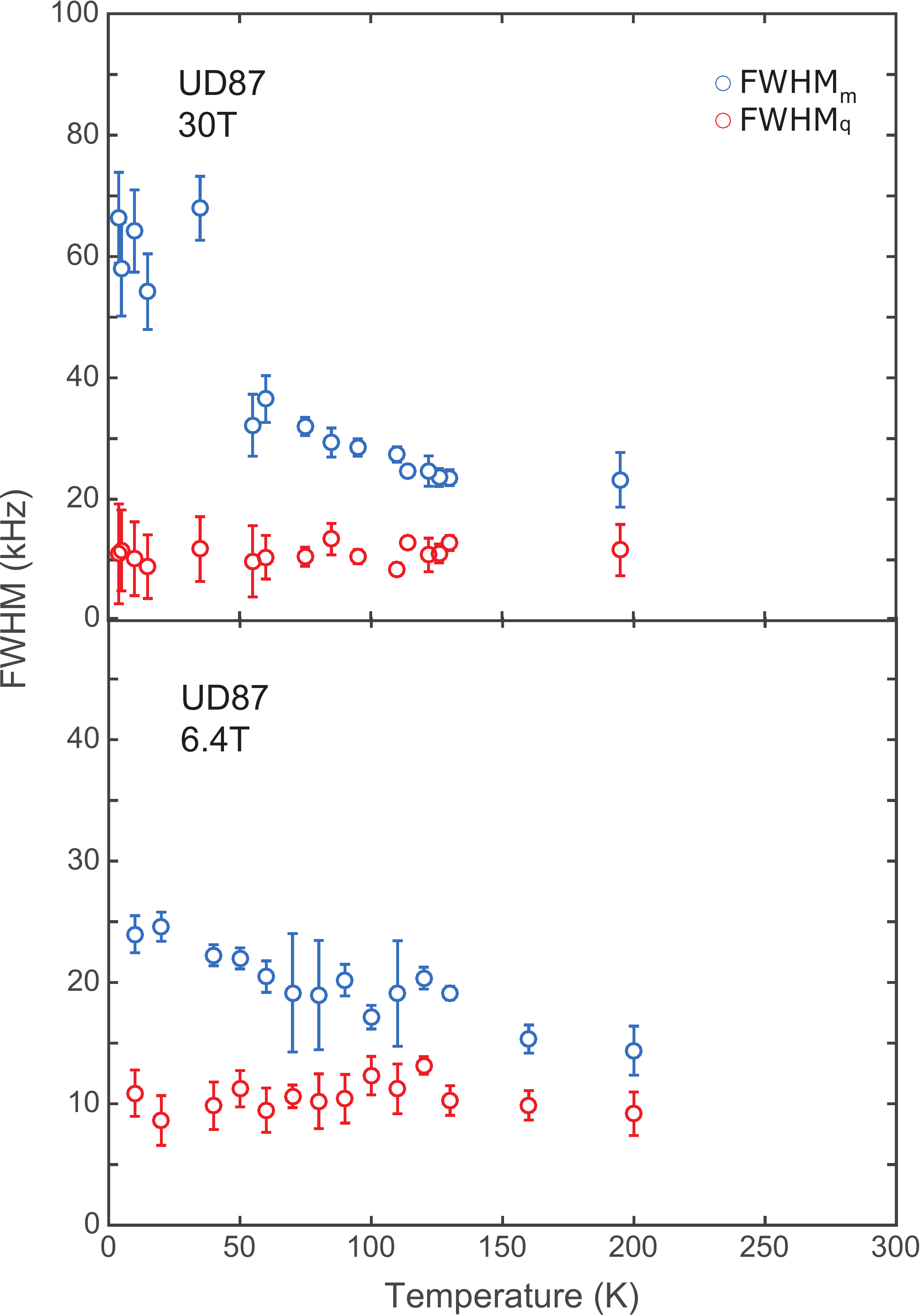}}
\caption {Magnetic and quadrupolar contributions to the total linewidth as a function of temperature  obtained from the fitting.  Error bars are statistical.}
\label{Lee.FigureS5}
\end{figure}

\begin{figure}[!ht]
\centerline{\includegraphics[width=10cm]{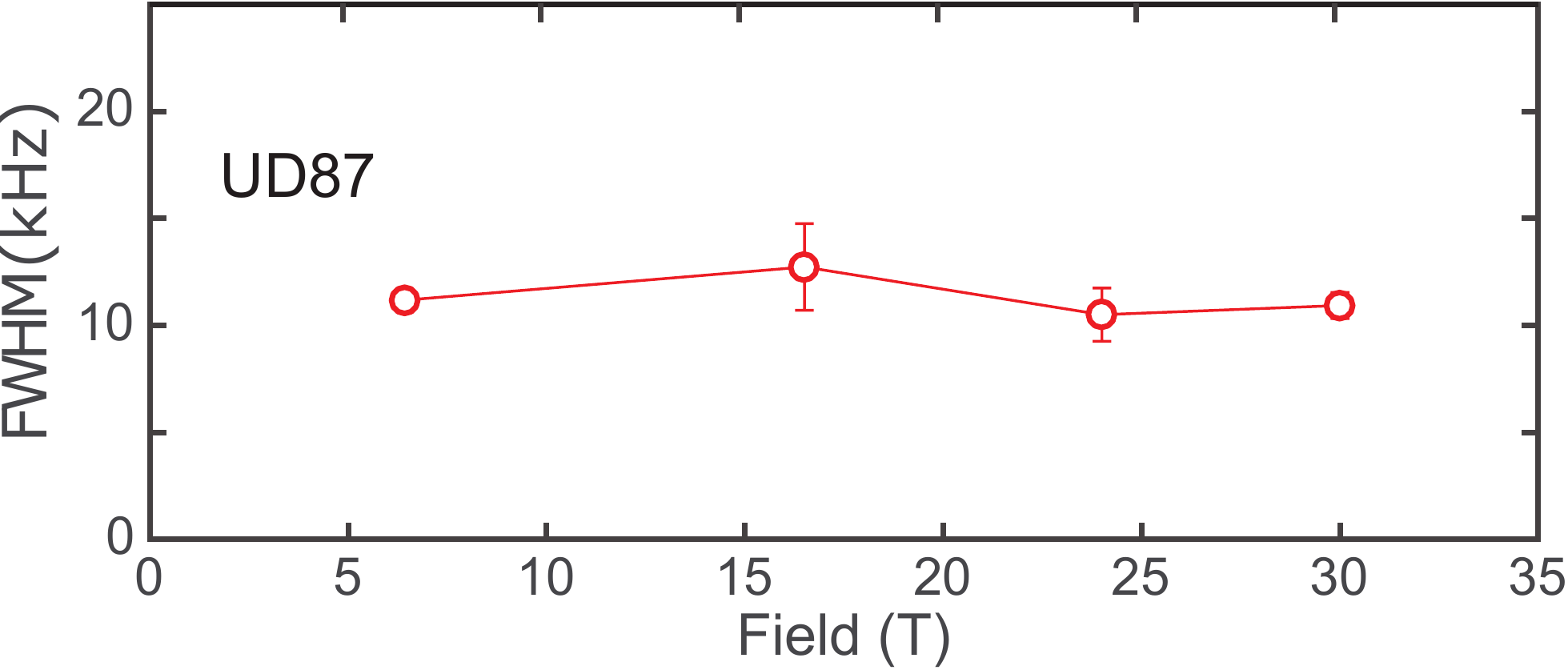}}
\caption {To get a linewidth that is representative of one field, we have averaged $\it{\sigma_q}$ at all temperatures at that field. The error bars shown here are statistical.}
\label{Lee.FigureS6}
\end{figure}

\begin{figure}[!ht]
\centerline{\includegraphics[width=10cm]{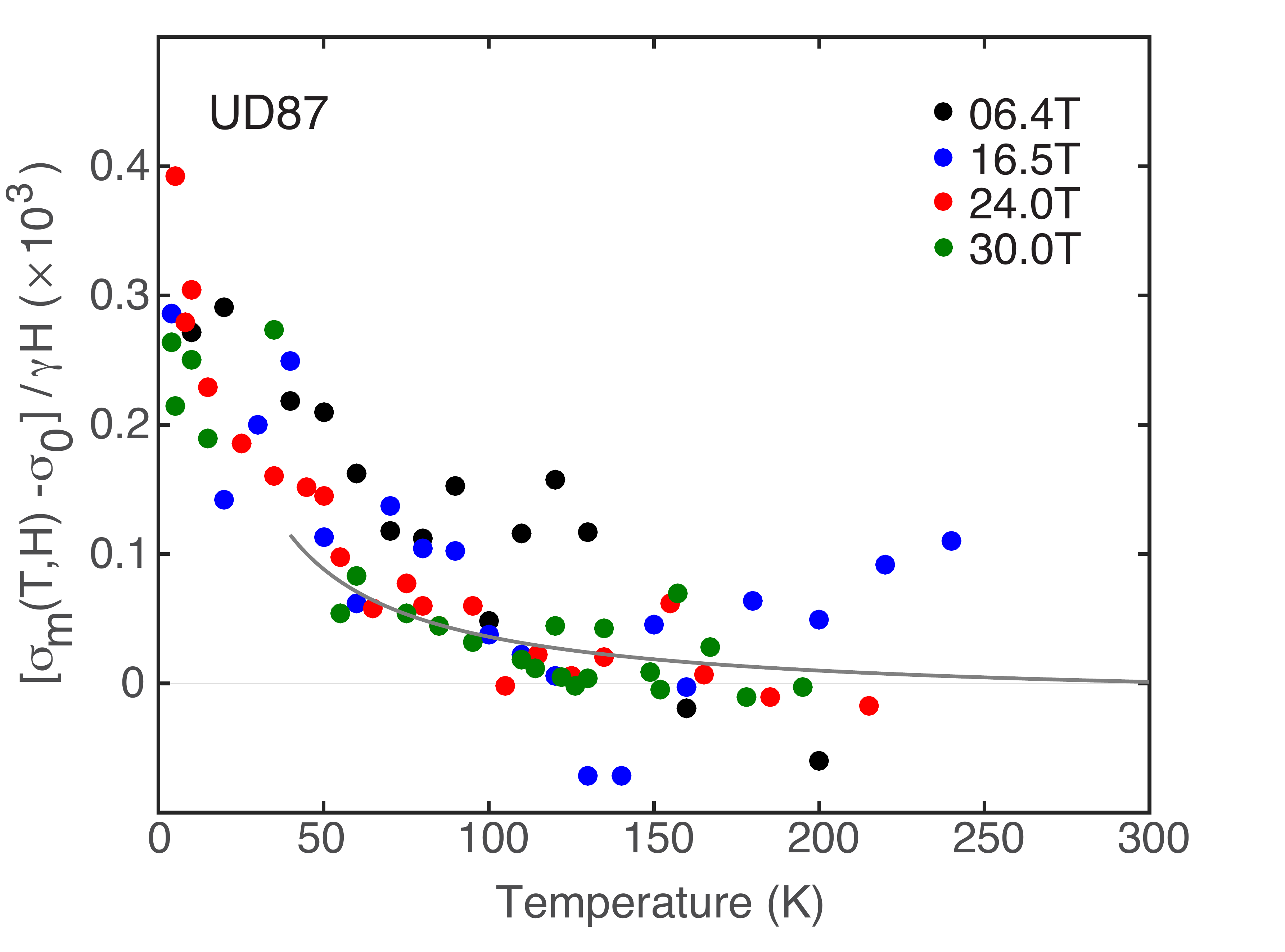}}
\caption {A single Curie-Weiss curve represents temperature and magnetic field dependence of the magnetic linewidth. In our underdoped samples the intrinsic linewidth is  $\sim 10$\,\,kHz, with the Curie-Weiss constant, $\it{C}$, and the Curie-Weiss temperature, $\Theta$, 30\,\,K/T and $\sim$ 0\,K, respectively. We have avoided fitting the magnetic contribution to the linewidth below the vortex freezing transition near $T=40$\,K.}
\label{Lee.FigureS7}
\end{figure}

\indent
\indent
There is a systematic behavior in the temperature dependence of linewidth components which emerges from the fitting analysis of  experimental spectra described in the text. Figures~\ref{Lee.FigureS5}, \ref{Lee.FigureS6}, and \ref{Lee.FigureS7} show  temperature and magnetic field dependence of the quadrupolar and magnetic linewidth contributions. The quadrupolar contribution was found to be temperature and magnetic field independent while the magnetic contribution increases with increasing field and decreasing temperature.

The field and temperature dependence of $\sigma_{m}$ for our two underdoped samples suggests Curie-Weiss behavior, shown in Fig.~\ref{Lee.FigureS5} and \ref{Lee.FigureS7}, where $\sigma_{m}$ is divided by, $\gamma H$, consistent with the equation, 

\begin{equation}
\label{eqS7}
\sigma_m(T,H) = \sigma_0 + \frac{C H}{T-\Theta}
\end{equation}

The magnetic contributions for different fields coalesce into a common temperature dependence, that can be understood in terms of paramagnetic impurities as reported for Bi2212~\cite{Chen2008} and Y123~\cite{Bobroff1997}.

\section{VI. Apical Oxygen}

\begin{figure}[!ht]
\centerline{\includegraphics[width=10cm]{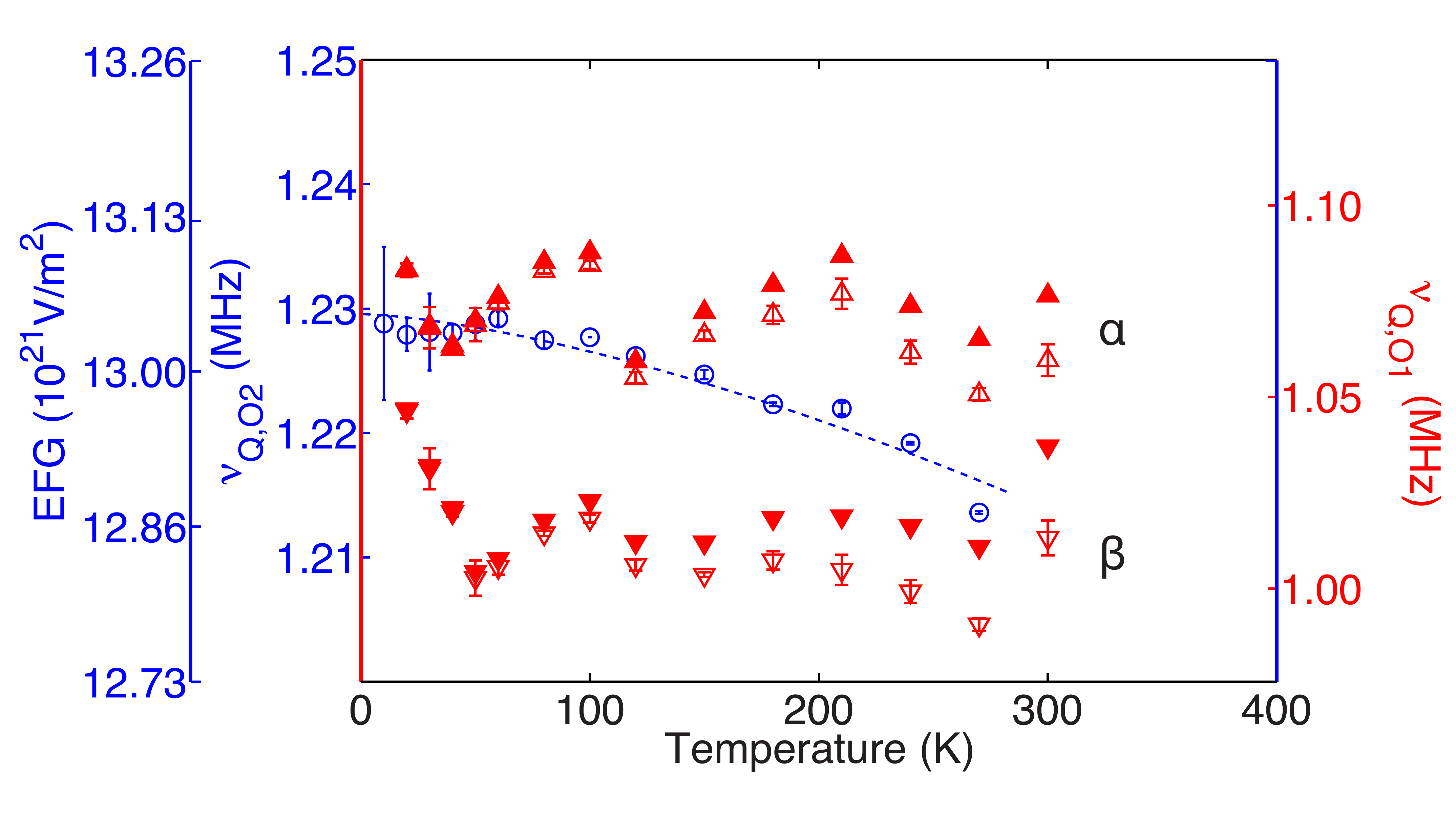}}
\caption {The phenomenological temperature dependence of lattice parameters for simple metals is shown by a dashed line. The origin of the decrease in $\it{\nu_{Q}}$ with increasing temperature at the apical site is therefore due to the temperature dependence of lattice parameters. The planar $\it{\nu_{Q}}$ for $\alpha$ and $\beta$ sites in the two-site model are shown after subtraction of this background temperature dependence. Solid (open) markers are after (before) the subtraction.  Error bars are statistical. } 
\label{Lee.FigureS8}
\end{figure}

Extending our earlier work~\cite{Mounce2013} we have also studied $\it{\nu_{Q}}$ of the apical oxygen at H = 30.0 T (Fig.~\ref{Lee.FigureS8}). The quadrupolar frequency is extracted from the quadrupolar separation of ($\pm\frac{3}{2}$,$\pm\frac{5}{2}$) satellite peaks. The average linewidth of each peak in the spectrum is $\sim$\,10 kHz.

\noindent
Our calculation of the crystal EFG at the apical site obtained from this analysis is consistent with first principle calculations for Hg1201~\cite{Araujo2000}.

For non-cubic simple metals, the thermal response of the lattice parameters follows the following phenomenological temperature dependence ~\cite{Jena1976,Christiansen1976,Kaufmann1979} which accounts for a contribution from the lattice vibrations due to phonon modes. 

\begin{equation}
\label{eqS9}
\nu_{Q} = \nu_{Q0} (1-bT^{3/2})
\end{equation}

\noindent
For the apical oxygen, we have observed continuously decreasing EFG at the apical site with increasing temperature (Fig.~\ref{Lee.FigureS8}) whose temperature dependence can be accounted for by this model with $\nu_{Q0}$ = 1.230(1) MHz and $b = 2.5(4) \times$10$^{-6}$ K$^{2/3}$ in our  UD79 sample. The thermal expansion of lattice parameters measured by neutron diffraction have a similar functional dependence~\cite{Huang1995}.

The background dependence was subtracted from the planar data leaving only on-site lattice contributions to the EFG. The onset of an upturn of $\it{\nu_{Q,\alpha}}$ and $\it{\nu_{Q,\beta}}$ at a low temperature around 40 K corresponds to the vortex freezing temperature of this sample (UD79) which was separately determined from NMR T$_{2}$ investigation of $^{63}$Cu and $^{17}$O nuclei~\cite{Lee2016}; however, the origin of this effect is unknown.

\bibliographystyle{naturemag}
\bibliography{ChargeOrder}